\documentclass[a4paper, twocolumn, floatfix, 11pt, accepted=2025-03-10]{quantumarticle}
\pdfoutput=1
\usepackage[utf8]{inputenc}
\usepackage{amsfonts}
\usepackage{physics}
\usepackage{graphicx}
\usepackage{overpic}
\usepackage{color}
\usepackage[dvipsnames]{xcolor}
\usepackage[hidelinks]{hyperref}
\usepackage{comment}

\usepackage[T1]{fontenc} 
\usepackage{enumitem}
\usepackage{booktabs}
\usepackage{float}
\usepackage{soul}

\usepackage{algorithm}
\usepackage{algpseudocode}

\usepackage{soul}

\definecolor{boxback}{HTML}{FFF8B5}
\definecolor{applegreen}{rgb}{0, 0.5, 0.0}
\definecolor{smoothred}{HTML}{C5232F}
\definecolor{mygreen}{rgb}{0,0.5,0}
\definecolor{myblue}{rgb}{0,0,0.75}
\definecolor{mymagenta}{cmyk}{0,1,0,0.12}

\usepackage{orcidlink}
\newcommand{\orciddaniel}{\orcidlink{0000-0001-7658-3546}}
\newcommand{\orcidpietro}{\orcidlink{0000-0001-5279-7064}}
\newcommand{\orcidsimone}{\orcidlink{0000-0002-8882-2169}}
\newcommand{\orcidmarco}{\orcidlink{0000-0002-3215-3453}}

\newcommand{\uulm}{Institute for Complex Quantum Systems, 
  Ulm University,
  Albert-Einstein-Allee 11, 89069 Ulm, Germany}
\newcommand{\unipd}{Dipartimento di Fisica e Astronomia "G. Galilei" \& Padua Quantum Technologies Research Center,
 Universit{\`a} degli Studi di Padova, Italy I-35131, Padova, Italy}

\newcommand{\pdinfn}{INFN, Sezione di Padova, via Marzolo 8, I-35131,
  Padova, Italy}

\newcommand{\nsuper}{n_{\rm \scriptscriptstyle S}}  
\newcommand{\nsamples}{n_{\rm r}}  

\newcommand{\opes}[1]{OPES} 
\newcommand{\Opes}[1]{Optimal tensor network sampling} 
\newcommand{\opesfull}[1]{optimal tensor network sampling} 
\newcommand{\Opesfull}[1]{Optimal tensor network sampling} 
\newcommand{\OPESfull}[1]{Optimal tEnsor network Sampling} 

\begin{document}

\title{Optimal sampling of tensor networks targeting wave function's fast decaying tails}

\author{Marco Ballarin\orcidmarco}
\affiliation{\unipd}
\affiliation{\pdinfn}

\author{Pietro Silvi\orcidpietro}
\affiliation{\unipd}
\affiliation{\pdinfn}

\author{Simone Montangero\orcidsimone}
\affiliation{\unipd}
\affiliation{\pdinfn}
\affiliation{\uulm}

\author{Daniel Jaschke\orciddaniel}
\affiliation{\uulm}
\affiliation{\unipd}
\affiliation{\pdinfn}

\begin{abstract}
  We introduce an optimal strategy to sample quantum outcomes of local measurement strings for isometric tensor network states. Our method generates samples based on an exact cumulative bounding function, without prior knowledge, in the minimal amount of tensor network contractions. The algorithm avoids sample repetition and, thus, is efficient at sampling distribution with exponentially decaying tails. We illustrate the computational advantage provided by our optimal sampling method through various numerical examples, involving condensed matter, optimization problems, and quantum circuit scenarios. Theory predicts up to an exponential speedup reducing the scaling for sampling the
  space up to an accumulated unknown probability $\epsilon$ from $\mathcal{O}(\epsilon^{-1})$ to $\mathcal{O}(\log(\epsilon^{-1}))$ for a decaying probability distribution.
  We confirm this in practice with over one order of magnitude speedup or multiple orders improvement in the error depending on the application. Our sampling strategy extends beyond local observables, e.g., to quantum magic.
\end{abstract}

\maketitle


In the era of the early development of quantum processors and simulators~\cite{Monroe2002,Monroe2013,Saffman2018,Wendin2017,Bruzewicz2019,Adams2020,Kaushal2020,Kjaergaard2020,Morgado2021},
a high demand arises for numerical emulators of programmable quantum devices: classical algorithms capable of replicating the input, real-time processing, and ouput of a quantum machine at small and intermediate scales~\cite{nagy2019,qiskit2023,cuquantum2023,Efthymiou2022}. Such demand is motivated by benchmarking and certification of the programmable quantum device \cite{Jaschke2022} as well as exploring the extent of quantum advantage:
classical simulation of quantum machines has shown remarkable progress in recent experiments \cite{Arute2019,Zhong2020} and new numerical algorithms have been developed to overcome existing obstacles \cite{Feng2022,Feng2022.2,Liu2021,Bulmer2022,Daley2022,Oh2023}.

Tensor networks are often employed to numerically characterize intermediate-scale quantum systems and devices. As they store spatial correlations as compressed information, they are an invaluable tool for describing many-body quantum systems with limited entanglement content.
Various classes of tensor networks have been developed so far, each adapted to a specific correlation geometry: from matrix product states~(MPS)~\cite{White1992,Rommer1997,Rommer1999,Schollwoeck2005,Perez2007,McCulloch2007,Schollwoeck2011} via tree tensor networks~(TTN)~\cite{Shi2006,Tagliacozzo2009,Murg2010,Gerster2014} to projected entangled pair
states~(PEPS)~\cite{Verstraete2008,Orus2014}, and more~\cite{Zwolak2004,Verstraete2004,Rizzi2008,Cincio2008,McCulloch2008,Jordan2008,Evenbly2009,Evenbly2014,Phien2015,Werner2016,Silvi2017,montangero2018,Jaschke2019,Reinic2021,Arceci2022,Banuls2022,Okunishi2022}.
Also in the context of sampling, tensor network methods have emerged as a powerful tool: Indeed, tensor network allow efficient output data acquisition from quantum states and processes~\cite{gomez2022, yang2023, dymarsky2022}.
Sampling of tensor network states has also been used to determine quantities that are otherwise difficult to measure, such quantum state {\it magic}~\cite{Lami2023},
entanglement characterization~\cite{Notarnicola2021},
or to sample from distributions via quantum states~\cite{Wild2021}.
Despite considerable recent advancements~\cite{Han2018,Bravyi2022}, we identified a sampling strategy meant to outpower known methods.

In this work, we present an efficient sampling procedure, which we label \OPESfull{} (\opes{}), applicable to
any isometric tensor network state~\cite{Zalatel2020}. The algorithm itself falls into the category of inverse transform sampling algorithms~\cite{Niederreiter1992,Devroye2006,Givens2012}; as additional challenge in comparison to the inverse transform sampling, we have no initial knowledge of the cumulative probability distribution or its inverse, which even by the end is known only partially for the many-body quantum state. Our approach exploits the capability of isometric tensor networks to efficiently acquire exact conditional probabilities for partial outcome strings, and carefully avoids re-sampling of previously explored outcomes.
\Opes{} method is thus particularly suited to characterize peaked, non-flat distributions,
where we observe a speedup beyond one order of magnitude over the standard sampling. The speedup to be expected depends foremost on the probability distribution, then on the number of samples or coverage to be achieved.
To demonstrate the usefulness of the approach, we consider two popular classes of isometric tensor networks, namely MPS and TTN.
We identify a substantial speedup both for sampling a fixed number of states as well as for covering a target portion of the probability space; while the exact probability of non-sampled space is known at all times.
The \opes{} algorithm benefits from storing in memory all the partial tensor network contractions, which characterize the conditional probabilities, evaluated until that instant, to avoid repeating calculations.
We analyze the advantage offered by the \opes{} method
in five practical examples from different quantum applications, such as image
representation, condensed matter or measurement of quantum magic. The examples include pure and mixed states.
The developed code is open source and included in the \emph{Quantum TEA}~\cite{qtealeaves_0_5_12} library.

This manuscript is structured as follows. First, we focus on the idea underlying \opes{} in Sec.~\ref{sec:theory}, then on the
numerical implementations in Secs.~\ref{sec:standard} and \ref{sec:methods}. In Sec.~\ref{sec:results}, we introduce classes of different
quantum states, each targeting different applications;
we sample with the two methods from these states and analyze the speedup that one can expect in each application. 
Finally, we summarize our results in Sec.~\ref{sec:conclusion}. The appendices
contain the full
algorithm in App.~\ref{app:real_algorithm}, as well as additional implementation
strategies for tensor networks, see App.~\ref{app:proj_meas_tn}.

\section{Theory and standard sampling from tensor network states                \label{sec:theory}}
%
\begin{figure*}[p]
    \centering
    \includegraphics[width=\textwidth]{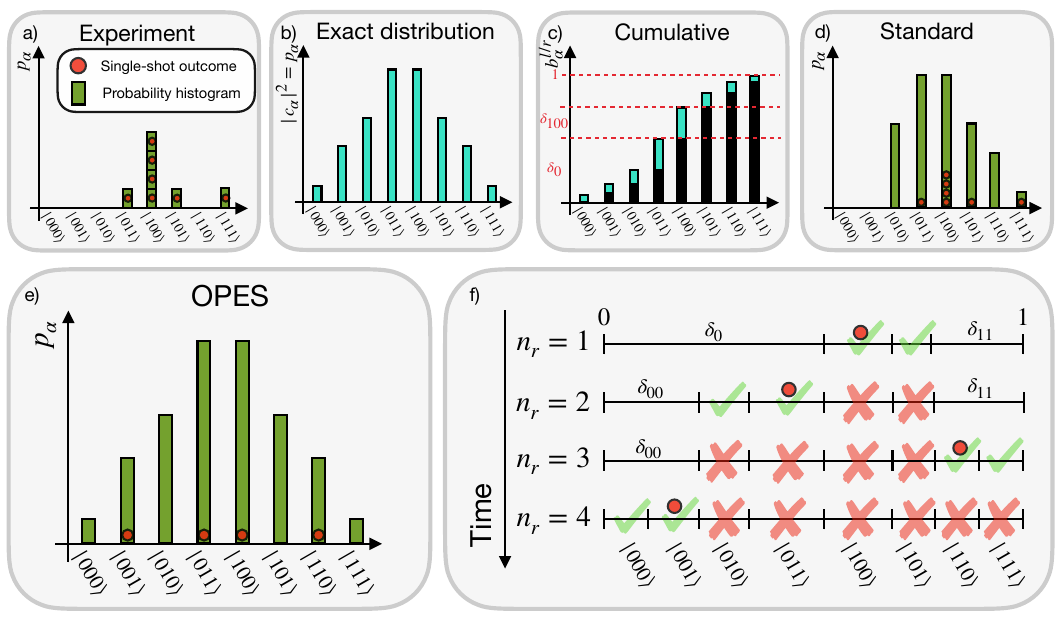}
    \caption{
    \emph{Different ways of sampling a given probability distribution function encoded in an $n=3$ qubit state.}
    a)~The motivation to sample by shots are experiments, where the probabilities are proportional to
    the normalized number of shots for a given state and converge in the limit of large numbers. Thus,
    the probability bar in green is proportional to the shots shown as red circles.
    b)~The actual probability distribution function is the reference distribution for any sampling method.
    c)~One can rewrite the exact probabilities into a cumulative probability function which highlights that each
    state $\alpha$ can be associated with an interval $\delta_{\alpha}$, see also Eq.~\eqref{eq:bound}, where the intervals for the subsystem-state
    $\ket{0}_{0}$ and state $\ket{100}$ are indicated as $\delta_{0}$ and $\delta_{100}$. The intervals are defined
    via the left and right boundaries of the intervals, i.e., $\delta_{\alpha} = (b_{\alpha}^{l}, b_{\alpha}^{r})$.
    The probability stemming from the previous states in the order is shown in black, the probability of the
    state itself is shown in cyan.
    d)~The standard sampling follows a shot-by-shot approach to sample states $\alpha$, but has access to the exact probabilities
    $p_{\alpha}$. Additionally, probabilities of states differing only in the last qubit of the bitstring are known
    at no additional cost, see non-zero bars without shot (red circle). The probability of the
    green bars do not sum to one, which accounts for the unknown probability space.
    e)~Using \opes{}, we avoid resampling and thus all shots are for different states; in detail,
    red circles fall into different bars. Moreover, pairs of probabilities are
    discovered together analog to \emph{d)} and even the substates of the
    the first $(n - 1)$ qubits of the shots in red are unique for each shot. Four steps are sufficient to sample a space of eight states, which
    generalizes to an upper bound of $\mathcal{O}(2^{n - 1})$ samples to scan a full Hilbert space. The
    distribution of the shots sampled in this way does not converge to the true
    distribution anymore as there is either zero or one shot per bin; one always uses the exact probabilities.
    f)~The \opes{} scaling is achieved by removing intervals from the space where the algorithm is
    allowed to sample; only previously not sampled states remain as potential next samples.
    While running the iterations top-down in this illustration, we gain partial knowledge of the cumulative probability
    distribution of \emph{b)}, i.e., of the intervals of the full state when reaching the
    green checkmark as well as subsystem intervals as indicated. For example in step
    $\nsamples = 1$, we identify the intervals $\delta_{0}$ and $\delta_{11}$ for subsystems. The intervals with green checkmarks are then removed for all
    the following iterations, see red crosses.
    \label{fig:graphical_abstract}
    }
\end{figure*}
%

Although we compare exclusively the performance of the standard algorithm and
the \opes{} algorithm for tensor networks in the following,
the motivation originates in experimental realizations and how sampling works there.
Figure~\ref{fig:graphical_abstract}a illustrates how experiments sample from
the underlying probability distribution by running multiple shots for three qubits. The ratio of the number of
shots for a state $\alpha$ over the total number of shots converges to the exact distribution in the
limit of large number of shots. Due to the central limit theorem~\cite{fischer2011}, the histogram
approximates the actual probability distribution. The basis states $\alpha$ forming the projection in
$z$-direction are an intuitive choice for the experiment, which for example can be sampled in an
atom-qubit experiment with a fluorescence snapshot~\cite{Monroe2002,Saffman2018,Morgado2021}. Similarly to the experimental results,
both tensor network sampling methods can generate the number of shots per state $\alpha$.

For simulations and theory, we encounter more possiblities to work with the actual
underlying probability distribution of the quantum state, i.e., depicted in
Fig.~\ref{fig:graphical_abstract}b.
While sampling generates the outcome of simultaneous measured local observables of qudits,
we focus on qubits as a building block for the many-body wave function and
a projection onto the Pauli operator $\sigma^{z}$ without loss of generality.
The possible measurement outcomes on a single qubit are its logical states
$\alpha \in \{ \ket{0}, \ket{1} \}$.
To build up the notation for the many-body case with a tensor network
state, we consider first probabilities arising from the wave function $\ket{\psi}$ for
$n$ qubits, i.e.,
\begin{align}                                                                                     \label{eq:vectorpure}
    \ket{\psi} &= \sum_{q_0, \dots, q_{n-1}=0}^1 c_{q_0,\dots, q_{n-1}}\ket{q_0\dots q_{n-1}}\nonumber\\
     &= \sum_{\alpha=0}^{2^{n}-1} c_{\alpha}\ket{\alpha} \, .
\end{align}
The sampling algorithms aim to reproduce the probabilities $p_{\alpha} = \left|c_{\alpha}\right|^{2}$.
We label the basis states $\alpha$ either with the binary number
$q_0\dots q_{n-1}$ or its integer representation. While we focus on a pure
state in Eq.~\eqref{eq:vectorpure}, the concept evidently holds as
well for mixed states. The order of the basis states $\alpha$ on the $x$-axis
in Fig.~\ref{fig:graphical_abstract}b is used throughout the manuscript.

Based on this given order, the cumulative probability distribution is generated and shown
in Fig.~\ref{fig:graphical_abstract}c. We obtain the ordered probability intervals, where
we define the probability $p_{\alpha} = \left|c_{\alpha}\right|^{2}$ as the width of
an interval $\delta_{\alpha}$ as
\begin{align}
  \delta_{\alpha} &= \left( b^l_\alpha, b^l_{\alpha} + |c_{\alpha}|^2\right), \label{eq:bound}\\
  b^l_\alpha &=\sum_{\beta<\alpha} |c_\beta|^2 \,. \label{eq:left_boundary}
\end{align}

The boundaries of each $\delta_{\alpha}$ are well defined within the complete
probability interval $(0, 1)$.  Thus, the left boundary $b^l_\alpha$ of the probability
interval associated with the state $\alpha$ can be defined as
reported in Eq.~\eqref{eq:left_boundary}. We stress that, as shown in Eq.~\eqref{eq:bound},
the right boundary is just $b^r_{\alpha}=b^l_{\alpha} + |c_{\alpha}|^2$. Inverse transform
sampling algorithms can be used if these intervals are all 
known~\cite{Niederreiter1992,Devroye2006,Givens2012}. Then, one can
sample a uniform random number in the interval $(0, 1)$,
which serves as value on the $y$-axis of Fig.~\ref{fig:graphical_abstract}c, and finds the
corresponding state $\alpha$ on the $x$-axis via the inverse cumulative probability
distribution. While the inverse cumulative probability distribution is easily accessible
once all $p_{\alpha}$ are known, they are initially unknown in the tensor
network scenario and the number of states
$\alpha$ grows exponentially in system size for many-body systems. But we can already anticipate based on Eq.~\eqref{eq:left_boundary} that
neither all $\{|c_\beta|^2\}_{\beta\neq\alpha}$ nor even $\{|c_\beta|^2\}_{\beta < \alpha}$
need to be known. Partial knowledge can be sufficient to attribute a random number to the corresponding interval and its state $\alpha$, e.g., knowing the Eq.~\eqref{eq:left_boundary}
for an intermediate state with less than $n$ qubits, i.e., the intervals $\delta_{\alpha'}$ with $\alpha'$ being
a subsystem of $\alpha$ containing the first $m < n$ qubits; we explain the details in Sec.~\ref{sec:methods}.

The standard approach to sample outcomes $\left(|c_{\alpha}|^2, \alpha\right)$
is the generation of bitstrings shot by shot. For each shot, the standard method consecutively
projects the qubits into their basis state~\cite{NielsenChuang}. The method can also be used directly on the level of observables~\cite{ferris2012}.
As we calculate probabilities
shot-by-shot, we obtain the shot count for each bitstring $\alpha$ at the end of the standard
sampling as shown in Fig.~\ref{fig:graphical_abstract}d. Instead of calculating the probability
as ratio of shot count over number of shots, tensor networks have direct access to a pair of probabilities $p_{\alpha}$
with each shot.
The pair always differs only in the measurement outcome of the last qubit.
Measuring the system qubit by qubit, we accumulate a product of conditional probabilities
\begin{align}                                                                   \label{eq:condprob}
    |c_{\alpha}|^2 =p(q_{n-1}|q_{n-2}\dots q_1 q_0) \dots p(q_1|q_0)p(q_0),
\end{align}
i.e., the easily accessible information about the couple $\left(|c_{\alpha}|^2, \alpha\right)$. The probability
for only the last qubit being in the other state depends on the same
probabilities of the state $\ket{q_{n-2}\dots q_1 q_0}$ and is obtained for free.
Therefore, Fig.~\ref{fig:graphical_abstract}d
shows the distribution of shots, the exact probability for all states measured via
shots from Eq.~\eqref{eq:condprob}, as well as the exact probabilities of some states
without shots. The latter appear due to differing only in the last qubit from a
state known due to a shot, e.g., $\ket{010}$ being obtained together with $\ket{011}$.

We observe that the standard sampling in Fig.~\ref{fig:graphical_abstract}d repeats the sampling
of the same bitstring multiple times, while the exact probabilities already known,
see state $\ket{100}$. We
summarize the problems in the standard sampling approach that the \opes{} method overcomes as follows:
\begin{enumerate}
    \item As the trial bitstrings $\ket{q_0\dots q_{n-1}}$ via the tensor
      tensor network measurement yield the probability from
      the exact distribution upon the first encounter, each of them might
      be sampled repeatedly without gaining any additional information;
    \item Similarly, it is statistically unlikely to sample marginal outcomes,
    thus the exponentially small tails of peaked distribution either remain largely
    unexplored or need an exponential time to be sampled.
\end{enumerate}
The fundamental idea of \opes{} is highlighted in 
Fig.~\ref{fig:graphical_abstract}e and f; details of \opes{} follow in
Sec.~\ref{sec:methods}.
Figure~\ref{fig:graphical_abstract}e shows the complete knowledge of all
probabilities after four shots. Because we can gain the probability of two states
with each measurement of a tensor network, the optimal number
of necessary measurements scales linearly with the size of the Hilbert space, i.e.,
as $\mathcal{O}(2^{n - 1})$. To reach this scaling, we permanently remove the
sampled outcome strings from the probability measure space in a way that they cannot
be sampled again, as shown in Fig.~\ref{fig:graphical_abstract}f. After each
step, we thus remove two intervals. Effectively, we renormalize the sampling
probabilities of the remaining, missing outcomes. Thus, the sampling probability
of a not yet sampled string increases at each shot. Let us suppose we just sampled the 
couple $(|c_{\alpha^*}|^2, \alpha^*)$ with probability $p_{\alpha^*}$ as the first
sample. Then, the probabilities of sampling any $\alpha\neq\alpha^*$ at the following
iteration are
\begin{align}                                                                         \label{eq:rescale_p}
    p_{\alpha\neq\alpha^*}\rightarrow\frac{1}{1-p_{\alpha^*}}p_{\alpha\neq\alpha^*}>p_{\alpha\neq\alpha^*} \, ,
\end{align}
and so forth.
Thus, \opes{} increases the probability to draw a bitstring at every iteration which results
in a faster converging strategy.

We observe an exponential speedup
reducing the scaling for sampling from $\mathcal{O}(\epsilon^{-1})$ to
$\mathcal{O}(\log(\epsilon^{-1}))$ for a decaying probability distribution, with
$\epsilon^{-1}$ being the cumulative unknown probability. For details, see
Sec.~\ref{sec:theo_scaling}.

\section{Standard tensor network sampling                                        \label{sec:standard}}

We briefly summarize the algorithm for the standard tensor network sampling, which
generates a single shot for two qubits as shown in Fig.~\ref{fig:sampling_schemes}a.
At the end of the algorithm, we have obtained the conditional probabilities of
Eq.~\eqref{eq:condprob}.
First, we draw a vector of random number$\vec{u}=\left(u_0, u_1, \dots, u_{n-1} \right)\sim U([0,1]), \; \vec{u}\in\mathbb{R}^{n}$
uniformly distributed in the interval $[0, 1]$. Then, we compute the probability $p_0$ of
measuring $\ket{0}$ on the qubit $0$, e.g., via the reduced density matrix $\rho_{0}$. Next,
we project $q_0$ of the state $\ket{\psi}$ in the measured subspace based on the random number $u_0$,
\begin{align}
    q_0 = \begin{cases}
    \ket{0} & \text{if } u_0<p_0 \, , \\
    \ket{1} & \text{if } u_0\geq p_0 \, .
    \end{cases}
\end{align}
Afterward, we repeat the procedure for the second qubit $q_{1}$ with the random number $u_1$
computing the conditional probability $p(q_1=\ket{0}|q_0=\ket{0})=p_{0|0}$ of
measuring $q_1$ in the state $\ket{0}$ conditioned on the measurement on $q_0$. In the
two-qubit example of Fig.~\ref{fig:sampling_schemes}a, we have all the information to
calculate the probability of the sampled state $p_{\ket{01}}$; moreover, we know the
probability $p_{\ket{11}} = p_{1|1} p_{0}$ which depends on the same sequence of projections
on the previous qubit, or qubits in the general case.
We repeat this procedure until the end of the qubit chain, obtaining the probability
of the state $\alpha$ as already pointed out in Eq.~\eqref{eq:condprob}.  For a
detailed description of how these quantities are efficiently computed via reduced
density matrices in an isometric tensor network, refer to App.~\ref{app:proj_meas_tn}.

\section{\Opesfull{}                                                               \label{sec:methods}}
%
\begin{figure*}[t]
  \begin{center}
     \includegraphics[width=0.8\textwidth]{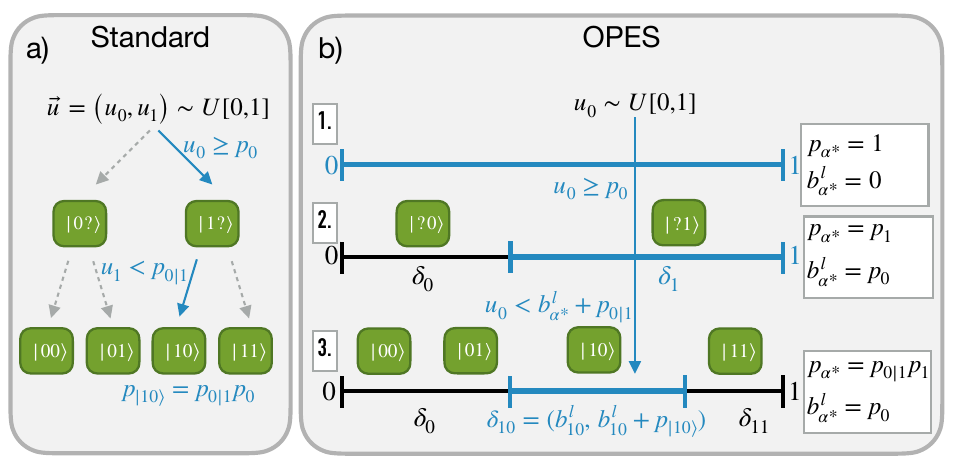}
    \caption{
      \emph{Projective measurements per qubit versus \opes{} sampling for one sample and two qubits.}
      We represent with $|?\rangle$ the unknown state a qubit, with $p_i$ the probability of measuring $\ket{i}$ on qubit $0$ and with  $p_{0|1}$ the probability of measuring $\ket{0}$ on qubit $1$ conditioned by the measurement on qubit $0$.
      a)~We show the standard sampling, where we determine the measurement results by sampling a random number for each qubit. The resulting information is just the probability of the final state. We use
      the random numbers top-down, first to decide on the state of
      the first qubit following the blue arrow to the left; then, 
      the random number for the second qubit samples state 0 and
      we follow the blue arrow to the right to the complete state
      $\ket{10}$. The reduced density matrix for the second qubits
      is calculated after using the projector $\ket{1} \bra{1}_{1}$
      on the first qubit in this example.
      b)~We report \opes{}, where a single random number is sampled for all the qubits. We collect knowledge of the probability intervals starting
      from an empty bitstring and the interval $(0, 1)$ containing
      the random number $u_{0}$, see step \emph{1)}. In step \emph{2)},
      we obtain the intervals $\delta_{0}$ and $\delta_{1}$ based on
      the reduced density matrix of the first qubit. We apply
      the projector $\ket{1} \bra{1}_{1}$ because
      $u_{0} \in \delta_{1}$ as highlighted in blue. We continue
      with the second qubit.
      By the end, we know the probability intervals $\delta_{10},\; \delta_{1}$ with $u_0\in\delta_{10}\in\delta_{1}$ while generating the bitstring, highlighted in blue;
      further, we know of $\delta_{0}$ and $\delta_{11}$.
      We enumerate the steps as in the main text, showing the evolution of the state probability $p_{\alpha^*}$ and its left boundary $b^l_{\alpha^*}$.
      }
      \label{fig:sampling_schemes}
  \end{center}
\end{figure*}

\Opes{} optimizes the sampling procedure measuring each state only once,
regardless of its probability. We divide the probability interval $[0,1]$
in unknown but ordered intervals $\{\delta_\alpha\}_{\alpha=0,\dots,2^n-1}$
following the idea of the cumulative probability distribution shown in
Fig.~\ref{fig:graphical_abstract}c. Considering one measurement as presented
in Fig.~\ref{fig:sampling_schemes}b, we highlight the two intervals $\delta_{10}$
and $\delta_{11}$ gained for the full state as well as the interval of the partial
state $\delta_{0}$.
We recall that the index $\alpha$ is the integer representation of the bitstring number.
After generating the sample $\left(\delta_\alpha, \alpha\right)$ with \opes{}, we remove $\delta_\alpha$ from the measurable intervals, and in the next iteration, we can only
measure a new state $\beta\neq\alpha$. This exclusion of previously sampled states enables us to explore exponentially decaying tails
efficiently.

The pedagogical and simplified algorithm for measuring simultaneous observables for two-level systems, depicted in Fig.~\ref{fig:sampling_schemes}b, is as follows:
\begin{enumerate}
    \item Set the state $\alpha^{*}$ to the empty
        bitstring, i.e., meaning no measurement has taken place so far.
        Therefore, the interval to draw uniform random numbers from is
        $\Delta=[0, 1]$; set the initial left boundary of the probability
        interval $b^l_{\alpha^*}$ and the state probability $p_{\alpha^*}$ to
        \begin{align}
            b^l_{\alpha^*}= 0, \quad p_{\alpha^*}=1 \, .
        \end{align}
        Sample $u_0\sim U(\Delta)$, where $U(\Delta)$ is a uniform distribution
        over a given interval.
        
    \item Compute the probability $p_0=p(q_0=\ket{0})$ of measuring $\ket{0}$
      on site $0$ and project the wavefunction $\ket{\psi}$ on a given state
      according to its probability and the random number $u_0$:
        \begin{align}
            q_0 &= \begin{cases}
            \ket{0} & \text{if } u_0<p_0, \\
            \ket{1} & \text{if } u_0\geq p_0,
            \end{cases} \\
            \ket{\alpha} &= \ket{\alpha^{*},q_{0}} \, .
        \end{align}
        Update the left boundary and the state probability according to the measure:
        \begin{align}
            b^l_{\alpha} &= \begin{cases}
            b^l_{\alpha^*} & \text{if } u_0<p_0, \\
            b^l_{\alpha^*} + p_0 & \text{if } u_0\geq p_0
            \end{cases}\\
            p_{\alpha} &= p_{\alpha^*} \cdot p(q_0).
        \end{align}
    
    \item Moving to the next qubit $i$, set $\alpha^{*} = \alpha$; the
        bitstring $\alpha^{*}$ now contains the qubits up to
        the $(i-1)^{\mathrm{th}}$ qubit. Compute the conditional
        probability $p_{0|\alpha^{*}}=p(q_i=\ket{0}|\alpha^{*})$.
        Select the measured state according to $u$ as in the previous point
        and set $b^l_{\alpha}, p_{\alpha}$ accordingly, i.e.:
        \begin{align}
            q_i &= \begin{cases}
            \ket{0} & \text{if } u_0 < b^l_{\alpha^*} + p_{0|\alpha^{*}}, \\
            \ket{1} & \text{if } u_0 \geq b^l_{\alpha^*} + p_{0|\alpha^{*}},
            \end{cases} \\            \ket{\alpha} &= \ket{\alpha^{*},q_{i}} \, , \\
            b^l_{\alpha} &= \begin{cases}
            b^l_{\alpha^*} & \text{if } u_0 < b^l_{\alpha^*} + p_{0|\alpha^{*}} \, , \\
            b^l_{\alpha^*} + p_{0|\alpha^{*}} p_{\alpha^*} & \text{if } u_0 \geq b^l_{\alpha^*} + p_{0|\alpha^{*}} \, ,
            \end{cases}\\
            p_{\alpha} &= p_{\alpha^*} \cdot p(q_{i}|\alpha^{*}).
        \end{align}

    \item Repeat step \emph{3.} until all the sites of the system are measured.\\ Further details are
    shown in Figure~\ref{fig:sampling_schemes}b: we have access to the information of many nested intervals after measuring all the sites. For example, if the final result is the interval $\delta_{10}$ we have information about the following intervals:
    \begin{align}
        \delta_{01}\in\delta_{1}\in[0, 1], \; \delta_0, \delta_{11}.
    \end{align}
    We denote $\delta_i$ the interval where only the first qubit has been measured to simplify the notation.
    We stress that knowing $\delta_0$ is enough to compute $b^l_{\alpha^*}$, and we do not need to know $\delta_{00}, \delta_{01}$.    
    In general, the obtained results are:
    \begin{align}
    b^l_{\alpha} &= \sum_{i} p\left(\overline{q_i}|\{q_j\}_{j=0}^{i-1}\right)\theta(q_i-\frac{1}{2}), \\
    p_{\alpha} &=  p(q_{n-1}|q_{n-2}\dots q_0)\dots p(q_1|q_0)p(q_0),
    \end{align}
    where $\theta(x)$ is the Heavyside step function and $\overline{q_i}=|1-q_i|.$

    Furthermore, we highlight that at each iteration we actually compute two probability intervals, i.e., $\delta_{q_0q_1\dots q_{n-2}q_{n-1}}$
    and $\delta_{q_0q_1\dots q_{n-2}\overline{q_{n-1}}}$.

    \item Remove the measured probability interval from our sampling interval, i.e., $\Delta=\Delta\setminus \delta_{\alpha}$.
        Each new sampled number $u_j~\sim U(\Delta)$ explores a 
        different probability interval $\delta_{\beta\neq\alpha}$.

\end{enumerate}
This presentation of the algorithm is pedagogical for understanding
the procedure and generates one sample. Steps 1) to 5) represent
in the complete algorithm only a single superiteration with a
number of random numbers sampled $\nsamples=1$;
we report the complete algorithm in Appendix~\ref{app:real_algorithm}.
The main differences are that for efficient computation of the intervals $\delta_\alpha$,
multiple random numbers $\{u_j\}_{j=1, \dots, n_r}$ are drawn for each superiteration, repeating then $\nsuper$ superiterations.
While we do not obtain anymore one new state $\alpha$ for each random number sampled from $U(\Delta)$, we are able to cache
intermediate results to optimize the procedure; nonetheless, we are still avoiding additional computation for the same interval.
For more information about the caching strategy using the
information of the intervals, see Appendix~\ref{app:proj_meas_tn}.

As a rule of thumb, if you expect your state to have a peaked distribution with
small tails you should run the first superiteration with $\nsamples > 1$, and then many
iterations with a bigger $\nsamples \gg 1$ to explore the space. Instead, it is much
better to always use $\nsamples \gg 1$ if you expect a flat
distribution, since the overhead due to
the cached intermediate results might reduce the performances.

\section{Theoretical scalings}\label{sec:theo_scaling}

We proceed now to analyze the optimality of \opes{} and its theoretical scaling
in treatable scenarios, i.e., for a flat and a decaying distribution.

\subsection{Uniform distribution}
In a uniform distribution, the probability of each state is $p = 1 / D$ where
$D$ is the dimension of the Hilbert space. Say that we sample anyway and search
for some state $\ket{\psi}$, we know that the probability to have sampled $\ket{\psi}$
after $n$ iterations is
\begin{align}
    p(n) &= 1 - \left( 1 - \frac{1}{D} \right)^{n} = \frac{D^{n} - (D - 1)^{n}}{D^{n}} \, .
\end{align}
Instead with OPES, we have
\begin{align}
    p_{OPES}(n) &= 1 - \prod_{i=0}^{n-1} \left( 1 - \frac{1}{D - i} \right) = \frac{n}{D} \, ,
\end{align}
which reflects that after $n = D$ we reach the state in any scenario. We rewrite it as
ratio
\begin{align}
    \frac{p}{p_{OPES}} &= \frac{1}{n} \sum_{i=0}^{n-1} \frac{(-1)^{i} \begin{pmatrix} n \\ n - 1 - i \end{pmatrix}}{D^{i}} \, .
\end{align}
The ratio does not tell us much and is in the interval $[p(D), 1]$. We might have
to change the question as ask how many samples are required to find the state $\ket{\psi}$
with a probability $q$. Then, we have
\begin{align}
    n(p(n) \ge q) &= \frac{\log(1 - q)}{\log\left(1 - \frac{1}{D}\right)} \, , \\
    n_{OPES}(p_{OPES}(n) \ge q) &= q \cdot D \, .
\end{align}
The expression for standard sampling diverges for $q \to 1$. The question
now is how fast the function diverges. The divergence logarithmically with $q$.

\begin{figure}[t]
    \centering
    \vspace{0.5cm}
    \includegraphics[width=0.45\textwidth]{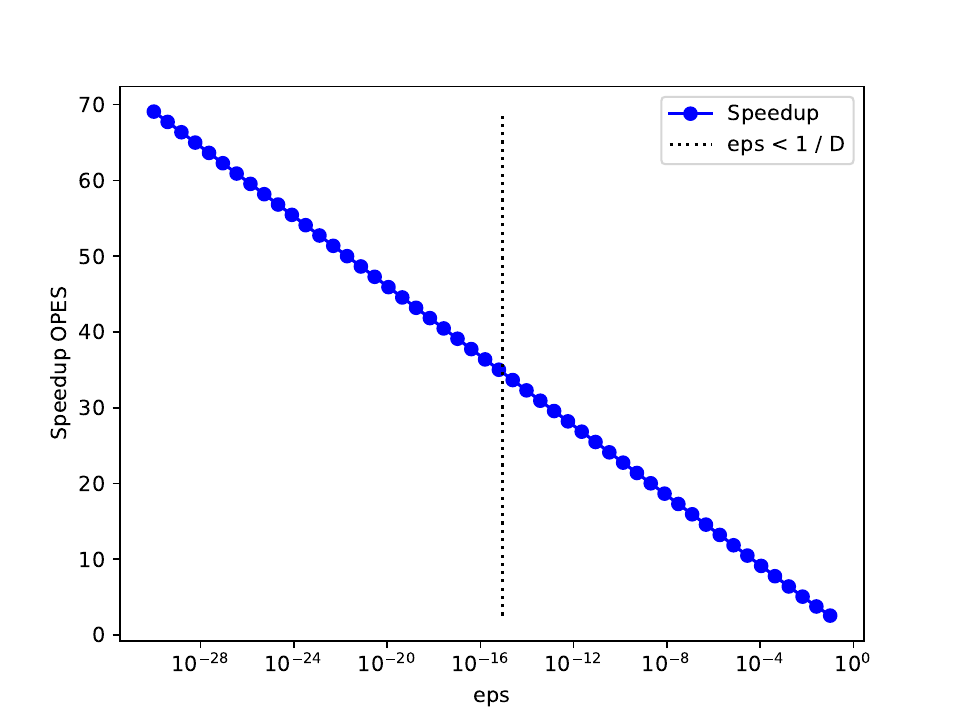}
    \caption{\emph{Establishing scaling for uniform distribution.} Speedup
    for OPES sampling all space up to $\epsilon$ for a uniform distribution
    and 50 qubits. We observe a log speedup with $\epsilon$.
    }
    \label{fig:uniform_scaling}
\end{figure}

To get an equal approach of the unsampled space $\epsilon$, we can analyze the Poisson
distribution $P_{\lambda = n / D}(k=0) = \lambda^{k} / k! \cdot e^{-\lambda}$
to get the number of states which have not been sampled, i.e.,
$D \cdot e^{- n / D}$:
\begin{align}
    n(\epsilon) &= -D \cdot \log(\epsilon) \, , \\
    n_{OPES}(\epsilon) &= D (1 - \epsilon) \, .
\end{align}
Assuming the overhead for bookkeeping is minimal, we have an OPES speedup
of
\begin{align}
    S(\epsilon) &= \frac{-\log(\epsilon)}{(1 - \epsilon)} \, .
\end{align}
The log-speedup of the uniform distribution can be seen as lower bound
for the speedop with OPES as it is the most difficult one to sample up
to an $\epsilon$.

\subsection{Exponentially decaying distribution}
We define the exponentially decaying distribution as follows:
\begin{align}                                                                    \label{eq:distexample}
    p_{i} = 2^{-i}, i = 1, \ldots 2^{n} - 1 \, , \qquad
    p_{i=2^{n}} = 2^{-(2^{n} - 1)} \, ,
\end{align}
where the last probability $p_{i=2^{n}}$ ensures normalization.
The number of qubits is $n$.

For the standard sampling, we assume that we have to sample probability
intervals of the order of $\epsilon$ to reach the corresponding coverage.
This statement is exact for the above example in Eq.~\eqref{eq:distexample}.
Then, the number of samples $j$ needed to find the outcome with probability
$\epsilon$ scales as
\begin{align}                                                              \label{eq:scaling_standard}
    \mathcal{O}_{\textrm{Standard}} &= \frac{1}{2} \mathcal{O}\left(\frac{1}{\epsilon} \right) \, ,
\end{align}
where the factor of one-half stems from the fact of obtaining two probabilities
from each sample.

Turning to the \opes{} algorithm, we want to avoid dealing with the
renormalized probabilities after removing intervals. Therefore, we assume
the samples are drawn according to their probability, highest probability
first. Then, we find the number
of probabilities to reach the given coverage, which is in case of
Eq.~\eqref{eq:distexample}
\begin{align}                                                              \label{eq:scaling_opes}
    \mathcal{O}_{\mathrm{\opes{}}} =& \frac{1}{2 \log(2)} \mathcal{O}\left(\log\left(\frac{1}{\epsilon}\right)\right) \, .
\end{align}
In this specific case of Eq.~\eqref{eq:distexample}, we can
also calculate the new probabilities after $j$ iterations, still
assuming we draw them in order
\begin{align}
    p_{i > 2j}^{j} &= 2^{2j} \cdot p_{i} \, ,
\end{align}
where the superscript indicates the iteration in the sampling.
Thus, the favorable scaling of \opes{} originates in the log-scaling
with $\epsilon^{-1}$ in comparison to the first-order polynomial scaling
with $\epsilon^{-1}$ for the standard sampling. We emphasize that this
arguments are independent of the size of the underlying Hilbert space.

\begin{figure}[t]
    \centering
    \includegraphics[width=\columnwidth]{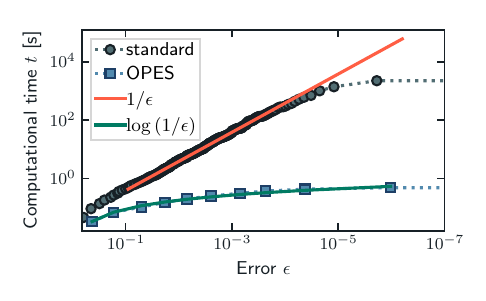}
    \caption{\emph{Computational scaling.} We show the computational time $t$ needed
       to obtain a given error $\epsilon$ for both methods. The probabilities
       are proportional to $p_{i} \propto d^{-i}$ with $d = 1.01$ for a system of
       $n=10$ qubits. The exponential speedup shows in the polynomial versus
       log scaling with $\epsilon$.}
    \label{fig:theoretical_scaling}
\end{figure}

To move forward and set up a numerical example, Eq.~\eqref{eq:distexample}
is decaying too fast to be captured by double precision. We generalize
the decaying probabilities to
\begin{align}                                                                    \label{eq:distexampled}
    p_{i} = \frac{d^{-i}}{\mathcal{N}} , i = 1, \ldots 2^{n} \, ,
\end{align}
where normalization is guaranteed by the scalar $\mathcal{N}$ and the
decay is parameterized by $d$.
In Fig.~\ref{fig:theoretical_scaling}, we report an experiment with
$n=10$ qubits and $d=1.01$, where we show that the numerical results
agree with the expected scalings. The scalings from Eqs.~\eqref{eq:scaling_standard}
and \eqref{eq:scaling_opes} are shown as additional lines for confirmation
with a suitable prefactor. If the threshold $\epsilon$ is too close
to the minimal probability $\min(p_{i})$ or below, the scaling breaks
down as expected.

The generalization to other discrete probability distributions is non-trivial.
For example, the discrete Gaussian distribution from Eq.~\eqref{eq:gaussianprofile}
can be considered for large system as continuous. Even then, we obtain
a scaling depending on the Gauss error function $\mathrm{erf}$. Moreover, an estimate if the distribution
has a smaller or larger benefit in comparison to Eq.~\eqref{eq:distexample} is not conclusive:
in case of the discrete Gaussian, both methods should have a better scaling
due to the smaller tails in the Gaussian versus Eq.~\eqref{eq:distexample}.
Therefore, a case-by-case analysis is necessary, either empirically as presented in Fig.~\ref{fig:gaussian},
by an analytic prediction of the scaling, or numerically constructing the scaling.

\section{Benchmarking sampling on applications                                                               \label{sec:results}}

In this section, we consider a variety of quantum states,
e.g., quantum images, quantum Ising ground states, Rydberg atoms, and quantum magic;
we compare the computational time and accuracy of the sampling methods.
Our results showcase the strengths of \opes{} over the standard
sampling for states with decaying tails in the probability distribution.
All computational times for the benchmarking of the sampling are referring
to Cineca's \emph{Galileo100} cluster, using a single compute node and 32 cores;
the node consists of Intel CascadeLake 8260 with 2xCPU x86 Intel Xeon Platinum 8276-8276L (2.4Ghz).

\subsection{Quantum state with Gaussian probability distribution with matrix product states}

\begin{figure}[t]
    \centering
    \begin{overpic}[width=0.95
    \columnwidth,unit=1mm]{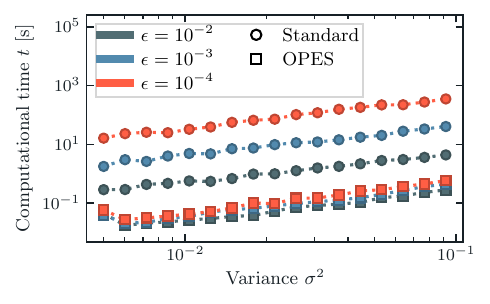}
        \put(-3, 60){a)}
    \end{overpic}
    \begin{overpic}[width=0.95 \columnwidth,unit=1mm]{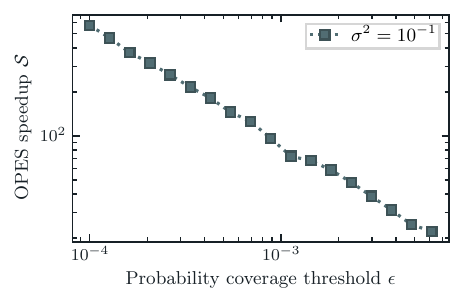}
        \put(-3, 60){b)}
    \end{overpic}
    \caption{\emph{Quantum state with a Gaussian probability distribution.}
        a)~The computational time $t$ of the standard sampling versus \opes{}
        for different probability coverage thresholds $\epsilon$ with $n=10$ qubits.
        States with a higher standard deviation $\sigma$ are more difficult to
        sample due to the tails of the distribution.
        b)~Computational speedup $\mathcal{S}$ of the \opes{} sampling over the standard one with respect to $\epsilon$ at fixed variance $\sigma^2$.
        }
    \label{fig:gaussian}
\end{figure}

First, we evaluate the performance of \opes{} sampling on a Gaussian quantum state.
The probability density of the state $\ket{\psi}$ has a Gaussian profile
\begin{align}                                                                    \label{eq:gaussianprofile}
    \ket{\psi} = \frac{1}{\mathcal{N}}\sum_{i=0}^{2^n-1} \sqrt{e^{-\frac{(x_i)^2}{2\sigma^2}}}\ket{i},\; x_i = \frac{i-\overline{i}}{2^{n}-1},
\end{align}
where the state is a superposition of basis states $\ket{i}$ that one samples.
We generate the state exactly and then convert it to an MPS representation without any truncation.
The parameter $n$ sets the dimension of the Hilbert space in terms of qubits;
the standard deviation $\sigma$ defines how close the state is to a delta-peak
and the value $\bar{i}$ shifts the maximum of the Gaussian between states. The
scalar $\mathcal{N}$ ensures that the state $\ket{\psi}$ is normalized.
Then, we compare the computational time $t$
required by two different methods to achieve a certain level of accuracy in characterizing the state distribution.
Specifically, we aim to know the state's distribution up to a given threshold~$\epsilon$.
Therefore, we define the coverage $\mathcal{C}$ and threshold $\epsilon$ of
a set of states with $\{ \delta_{1}, \ldots, \delta_{m} \}$ as
\begin{align}                                                                               \label{eq:coverage}
  \mathcal{C} &= \sum_{\alpha=1}^{m} \left| c_{\alpha} \right|^{2} \, , \qquad
  \epsilon = 1 - \mathcal{C} \, .
\end{align}
We show the results for ten qubits $n=10$, $\overline{i}=2^{n-1}-1$, and different values of the variance $\sigma^2$ in Fig.~\ref{fig:gaussian}.
The computational time for characterizing the probability interval up to the threshold $\epsilon$ increases for both methods as the variance increases. However, we find that \opes{} sampling
outperforms traditional sampling significantly due to its efficient use of cached information and the tail's sampling.
Let us define the speedup $\mathcal{S}$
\begin{align}
    \mathcal{S}=\frac{t_{\text{standard}}}{t_{\text{\opes{}}}},
\end{align}
with $t_{\text{\opes{}}}$ and $t_{\text{standard}}$ the computational time
for the \opes{} and standard sampling, respectively. In
Fig.~\ref{fig:gaussian}b, we show that the speedup achieved scales with
the threshold $\epsilon$, increasing from one order of magnitude
for $\epsilon=10^{-2}$ to three orders of magnitude for $\epsilon=10^{-4}$.
An in-depth argument on the scaling and speedup can be found in
Sec.~\ref{sec:theo_scaling}, but we have no expected theoretical scaling
for a comparison in case of Eq.~\eqref{eq:gaussianprofile}. Finally,
the computational time $t$ for \opes{} sampling is almost independent
of the chosen coverage threshold. These results demonstrate the
superiority of \opes{} sampling over traditional sampling when one aims
to cover most of the probability distribution.

\subsection{Quantum image with matrix  product states}

\begin{figure}[t]
    \centering
    \begin{overpic}[width=0.95 \columnwidth,unit=1mm]{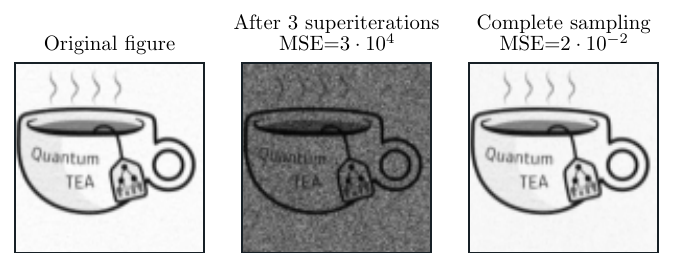}
        \put(-5, 33){a)}
    \end{overpic}
    \begin{overpic}[width=0.95 \columnwidth,unit=1mm]{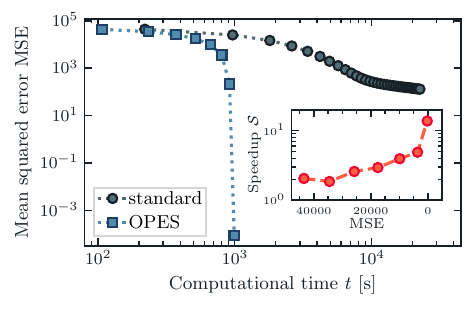}
        \put(-3, 60){b)}
    \end{overpic}
    \caption{\emph{Flexible representation of quantum images (FRQI) algorithm.}
      a)~Quantum tea logos used in the flexible representation of quantum images algorithm.
      From left to right, we show the original image, the reconstructed image after
      three superiterations of the \opes{} sampling, the image at the end of the \opes{} sampling.
      b)~Computational time $t$ of the standard sampling versus \opes{} for
      reaching a given mean square error MSE in the image reconstruction.
      In the inset, the speedup of \opes{} over the standard method is highlighted.
    \label{fig:quantum_image}}
\end{figure}

We proceed and analyze a state obtained after the application of a quantum circuit.
In particular, we are not interested in exploring final states of quantum algorithms,
since these algorithms are built to be easy to sample to minimize the number of
measurements; examples of such algorithms are the Grover or
Shor algorithms~\cite{grover1996, shor1994}. We instead investigate a state obtained
"at runtime". To perform operations on images using quantum hardware,
we first have to load the image. There are several possible ways of doing so: amplitude
encoding, angle encoding, or flexible representation of quantum images (FRQI)~\cite{le2011}. We focus on the latter case, encoding images
using the algorithm in Ref.~\cite{amankwah2022}. The question at hand is if we can reconstruct
the image with high accuracy after loading it, using tensor network sampling. This
procedure helps in understanding if the image is correctly reproduced even for
sizes that are not accessible with the exact state vector simulation. In particular, we choose to encode the quantum tea logo, shown in Fig.~\ref{fig:quantum_image}a. The logo is a $600\times600$ grey-scale image with a total of $M = 360000$ pixels encoded in a $20$-qubit state. We now proceed to a quantitative study of the sampling.
In Fig.~\ref{fig:quantum_image}b, we show the computational time $t$ needed to reach
a given mean square error MSE with the original image, defined as:
\begin{align}
    \text{MSE}(t) = \frac{1}{M}\sum_{i=1}^{M} \Big(\text{P}_{sv}^i - \text{P}_{sm}^i(t) \Big)^2,
\end{align}
where $\text{P}_{sv}$ is the picture obtained from the complete state vector of the state and
$\text{P}_{sm}(t)$ is the sampled picture at time $t$. The index $i$ runs over the pixels of the
image.
Thus, the MSE is improved by over two orders of magnitude in the figure of merit while
saving one order of magnitude in computational time. We also show the logos retrieved
from the quantum state at different superiterations of the \opes{} sampling in
Fig.~\ref{fig:quantum_image}a. The FRQI algorithm benefits in our opinion the most
from the \opes{} sampling out of our examples. As we see in the inline of
Fig.~\ref{fig:quantum_image}b, the speedup $\mathcal{S}$
of \opes{} over the standard method is exponential in the MSE, which is agreement
with Sec.~\ref{sec:theo_scaling}.

\subsection{Quantum Ising model with tree tensor networks}\label{sec:ising}

As the next example, we consider the quantum Ising model~\cite{SachdevQPT} as an example from
condensed matter. We use the definition of the quantum Ising model as
\begin{align}
    H &= J \left( -\sum_{i=1}^{n - 1} \sigma_{i}^{z} \sigma_{i+1}^{z} - g \sum_{i=1}^{n} \sigma_{i}^{x} \right) \, ,
\end{align}
where the parameter $J$ set the energy scale and $g$ tunes the
external field, respectively. The Pauli matrices acting on site $i$ are
$\sigma_{i}^{\alpha}$. From the two limits, we expect three regions with
respect to the sampling:
(i)~close to the ferromagnetic limit, the states almost align with our measurement basis;
(ii)~close to the paramagnetic limit, the local product states are orthogonal to the local measurement
basis; and (iii)~the region around the quantum critical point with high entanglement. The
latter two are thus potentially difficult to sample, especially in the case
of the paramagnetic limit where the probability of each state scales inversely
with the dimension of the Hilbert space.

\begin{figure}[!h]
  \begin{center}
    \vspace{0.1cm}
      \begin{overpic}[width=0.85 \columnwidth,unit=1mm]{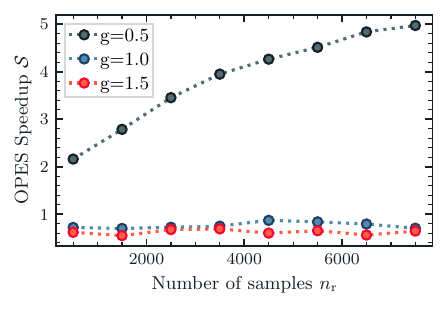}
        \put(-3, 60){a)}
      \end{overpic}
      \begin{overpic}[width=0.85 \columnwidth,unit=1mm]{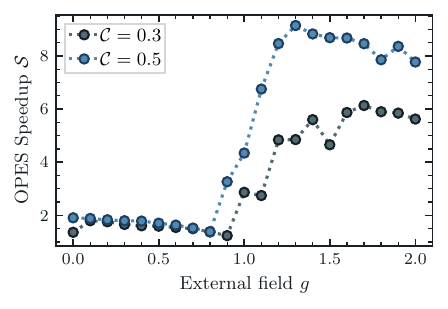}
        \put(-3, 55){b)}
      \end{overpic}
      \begin{overpic}[width=0.85 \columnwidth,unit=1mm]{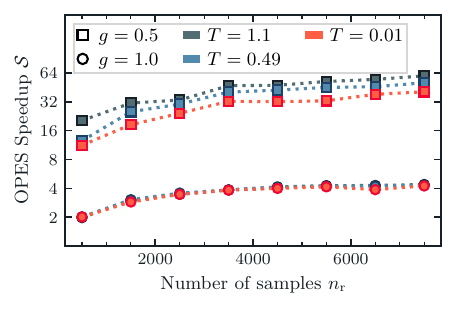}
        \put(-3, 60){c)}
      \end{overpic}
    \caption{\emph{Comparisons for the quantum Ising model and tree tensor networks.} 
       We calculate the speedup $\mathcal{S}$ of \opes{}
       as the ratio of the computational time of the standard method over \opes{}.
       a)~We sample $\nsamples$ states within one superiteration from the ground state
       at the critical point $g = g_{c} = 1.0$ for $64$ qubits.
       b)~The speedup $\mathcal{S}$ to reach a coverage $\mathcal{C}$ for 16 qubits
       across the phase diagram. The
       paramagnetic phase is difficult as the eigenstate does not align with
       the measurement basis.
       c)~The speedup $\mathcal{S}$ to produce $\nsamples$ samples
       within one superiteration for the density matrix at finite temperatures $T \in [1.1,0.49,0.01]$
       and two external fields $g$.
        \label{fig:ising}}
  \end{center}
\end{figure}

The sampling of a fixed number of samples $\nsamples$ within one superiteration
scales better either for the standard or \opes{}
approach according to Fig.~\ref{fig:ising}a, depending on the
quantum phase represented
by one point in the ferromagnetic and paramagnetic phase, see $g=0.5$ and $g=1.5$ respectively,
and an additional point close to the quantum critical point with $g=1$. We consider
a system size of $n = 64$. The probability distribution for $g=1$ and $g=1.5$ are
too similar to a uniform distribution and \opes{} does not show a benefit here, but is
even slower due to a potential overhead necessary for caching intermediate results.
If we aim to obtain a certain coverage $\mathcal{C}$ of the
probability space instead of a fixed number of iterations, the task of reaching the coverage becomes computationally more expensive, especially for flat-distributions. For example, we have to sample in an exponentially large space in the paramagnetic limit
because the local eigenstates
are orthogonal to our measurement basis. Keeping the measurement basis fixed throughout
helps to understand the scaling for other models where a change into the "good" measurement
basis might not be obvious. But we reduce the number of qubits
to $N =16$ and aim for coverages $\mathcal{C} = 30\%$ and $\mathcal{C} = 50\%$ in Fig.~\ref{fig:ising}b;
without recucing the number of qubits $N$, we need to sample $2^{N - 1}$ states in the
paramagnetic limit for $N = 64$ to reach $\mathcal{C} = 50\%$.
The better scaling of the \opes{} method for all fields $g$ originates from the fact that a
new iteration does not resample states from previous iterations. In contrast,
the standard method has a probability of almost 50\% to throw a state away
because the generated state is already known when reaching $50\%$ coverage.
Thus, the \opes{} still pays off in flat distributions for high
coverages $\mathcal{C}$, though this scenario is restricted to moderate
number of qubits due to the exponential growth of the Hilbert space.

The third example moves from pure states, which we consider throughout
all the other examples, to density matrices. We generate finite
temperature density matrices for the external fields $g=0.5$ and $g=1.0$ via
an imaginary time-dependent variational principle (TDVP) time evolution~\cite{Haegeman2016} with a locally purified tensor
network (LPTN)~\cite{Werner2016}, which is then converted into a tree tensor
operator (TTO)~\cite{Reinic2021}. The behavior for the external
field resembles the ground state, i.e., \opes{} shows a greater benefit inside the
ferromagnetic phase for the data points with $g=0.5$. Surprisingly, the data
indicates a bigger speedup at higher temperatures although the high-temperature
limit itself is the completely-mixed state with a flat distribution.
In summary, any problem where a lower limit on the unsampled space is required
can profit significantly from \opes{} even in scenarios similar to the condensed matter
case across all phases treated here.

\subsection{Crosstalk errors in Rydberg atoms with tree tensor networks}

\begin{figure}[t]
  \begin{center}
    \vspace{0.1cm}
      \begin{overpic}[width=0.9 \columnwidth,unit=1mm]{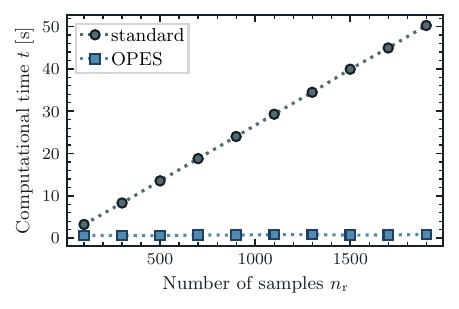}
        \put(-3, 60){a)}
      \end{overpic}
      \begin{overpic}[width=0.9 \columnwidth,unit=1mm]{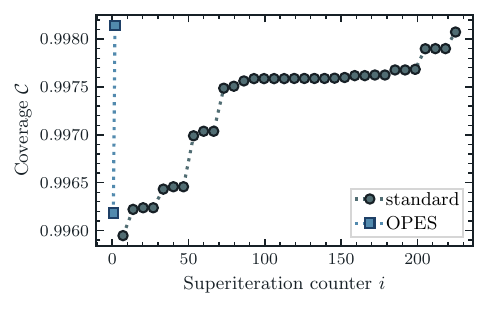}
        \put(-3, 55){b)}
      \end{overpic}
      \caption{\emph{Comparison for sampling an imperfect GHZ prepared under crosstalk.}
       The state is obtained from a simulation of Rydberg atoms
       encountering crosstalk during the preparation, which leads to deviations
       from the GHZ state.
       a)~We sample a fixed number of samples $\nsamples$ within one superiteration, where the major probability
       in the states $\ket{00\ldots 00}$ and $\ket{11\ldots 11}$ leads to the better
       scaling of \opes{}.
       b)~We aim for a coverage $\mathcal{C} \ge 99.8\%$ and show the coverage
       $\mathcal{C}$ during the superiterations of the algorithms.
        \label{fig:rydberg}}
  \end{center}
\end{figure}

An error analysis is useful for the benchmarking of quantum processing units,
e.g., for the state preparation. In such a scenario, we expect to be almost
perfectly in the target state, where the remaining probability represents the
error. Depending on the fidelity of the preparation, the measurement of a
state with error has a small probability. The preparation of the
Greenberger-Horne-Zeilinger state (GHZ) with errors including the projective
measurement is shown in Ref.~\cite{Jaschke2022} and we consider here the aspect
of the computational scaling of analyzing such a state.
Figure~\ref{fig:rydberg}a shows the computational time $t$ for an imperfect 64-qubit
GHZ state, where errors have been introduced during the protocol
due to crosstalk; while the computation time for projective measurements
scales with the number of samples for the standard approach, the \opes{}
approach has an almost constant computational time $t$. This almost constant time $t$
is related to the two large probabilities for the states $\ket{00 \ldots 00}$
and $\ket{11 \ldots 11}$. This scenario is one of the most favorable for \opes{}.

We further extract the coverage $\mathcal{C}$ as defined in Eq.~\eqref{eq:coverage} from the quantum state for each sampling method, i.e., how
much of the probability space has not been sampled. This data could be
used to generate data according to a quantum state's probability distribution up to a certain accuracy
even without a tensor network representation. Figure~\ref{fig:rydberg}b
shows how the coverage improves over time with the superiterations of the algorithm,
while we sample $\nsamples = 250$ states per superiteration for the default approach in comparison
to $\nsamples = 25$ states for \opes{}. As \opes{} can target exclusively
previously unexplored probability intervals, we observe a significant
advantage.

\subsection{Magic with matrix product states }

\begin{figure}[t]
    \centering
    \includegraphics{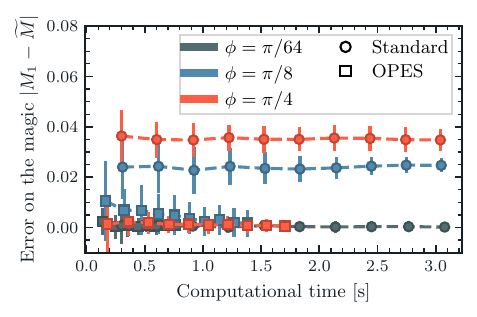}
    \caption{\textit{Magic estimation through sampling.}
    We compare the standard and \opes{} methods in computing the magic of the state known by its theory formula, see $\widetilde{\mathcal{M}}$ in Eq.~\eqref{eq:magic_state},
    for $n=10$ qubits and different parameters $\phi$ with the same number of samples.
    The points are obtained as an average over ten realizations.
    \Opes{} is always at least twice as fast as the standard method and converges faster
    to the true value $\mathcal{M}_{1}$ for $\phi=\frac{\pi}{8}$ and $ \phi= \frac{\pi}{4}$.
    }
    \label{fig:magic}
\end{figure}

To stress the generality of the method mentioned in Sec.~\ref{sec:theory},
we now generalize the sampling procedure to measure
the quantum magic of a quantum state, also known as non-stabilizerness.
Quantum magic is an important quantity since it limits the classical
simulation of quantum systems for stabilizer simulators~\cite{howard2014,seddon2021},
similarly to how the entanglement limits the simulations with tensor networks.
Quantum magic can be measured through Stabilizer Rényi Entropies (SREs)~\cite{leone2022, Tarabunga2023}.
Reference~\cite{Lami2023} introduces a technique to measure the SREs through standard sampling
of an MPS state. Thus,
applying \opes{} can speed up the process of measuring quantum magic.
The magic of a $n$-qubits state using the $m$-order SREs is defined as:
\begin{align}
    \mathcal{M}_m=\frac{1}{1-m}\log\left( \frac{1}{\mathcal{N}}\sum_{\mu=0}^\mathcal{N}\Pi(\vec{\sigma})^{m-1}\right)- n\log 2,
\end{align}
where $\Pi(\vec{\sigma})$ is the probability of measuring a given Pauli string, $\sigma_i\in\{\mathbb{I}, X, Y, Z\}$.
To apply \opes{} to measure $\mathcal{M}$, we target the $\Pi(\vec{\sigma})$, keeping into account
that each measure of probability on the single site $\pi(\sigma_i)$ branches into four possibilities
($\sigma_i\in\{\mathbb{I}, X, Y, Z\}$) instead of the two possibilities ($b_i = \{\ket{0}, \ket{1}\}$) treated up to now.
For an extensive explanation of how to measure $\pi(\sigma_i)$ efficiently with an MPS,
please refer to Ref.~\cite{Lami2023}.
As an example, we compute the order-1 SREs of the state
\begin{align}
    \ket{\psi(\phi)} = \left(\frac{\ket{0} +e^{i\phi}\ket{1}}{\sqrt{2}}\right)^{\otimes n}.
\end{align}
Notice that $\ket{\psi(0)}$ for $\phi = 0$ is a stabilizer state, and thus has zero magic.
The magic of the state is instead maximum for $\phi=\frac{\pi}{4}$. We call the exact 1-SREs of this state $\widetilde{\mathcal{M}}$:
\begin{align}
    \widetilde{\mathcal{M}} =& -\cos^2\phi\log\left(\left|\cos\phi\right|\right) \nonumber\\
    &-\sin^2\phi\log\left(\left|\sin\phi\right|\right).\label{eq:magic_state}
\end{align}
In Fig.~\ref{fig:magic}, we address the computation of $\mathcal{M}_1$ for $n=10$ qubits
using for \opes{} $n_S=10$ superiterations with $n_r=100$ samples each. The results are mediated over $10$
realization of the computations. We use for the standard method the same number of samples.
We notice that, due to the flatness of the distribution, \opes{} gives no relevant improvement when
$\phi=\frac{\pi}{64}\sim 0$. This result is to be expected, as already discussed for the paramagnetic phase
of the quantum Ising model in Sec.~\ref{sec:ising}. However, we can observe both an improvement in
the estimation of the magic and a speedup in the other two cases, i.e., $\phi=\frac{\pi}{8}$ and $\phi = \frac{\pi}{4}$.
\Opes{} is twice as fast as the standard method, and the estimation
of the quantum magic converges much faster with respect to the standard method.

\section{Conclusions                                                            \label{sec:conclusion}}

Overall, we have introduced the \opesfull{} (\opes{}), a sampling method for tensor network states
that avoids sampling known intervals and caches information for optimal performance.
From the computational statistics point of view, the algorithm represents an
extension to the inverse transform sampling, which is used to sample pseudo-random
numbers from an arbitrary probability distribution function by knowing its inverse
probability distribution function. The extension consists in the fact that we
never fully know the cumulative distribution function or its inverse of the quantum
state represented as a tensor network. The method is applicable
to all isometric tensor network states as well as exact diagonalization. We have investigated
several quantum states and compared the performance of the standard and
\opes{} methods in characterizing them; these results are obtained
for matrix product states~(MPS), tree tensor networks~(TTN), and tree tensor operators~(TTO). We compare to the standard sampling
method. An enhanced standard sampling, as available in Ref.~\cite{gray2018quimb}, can share some of the features
of OPES as the speedup at the cost of sampling as well as storing more random
numbers and loosing convenience in the numerical implementation.

In detail, we have shown that \opes{} offers a significant speedup compared
to the standard method, especially when covering most of the probability distribution is demanded.
On the one hand, this speedup is due to the efficient caching of
previous information employing the MPS or TTN; on the other hand, we work
with the underlying probability intervals of the full Hilbert space which
allow us to exclude previously explored intervals from future superiterations.
For example, \opes{} leads to a speedup of more than an order of
magnitude while achieving lower errors versus the standard sampling 
for applications as the flexible representation of quantum images (FRQI)
algorithm.
We have shown how the approach is generalizable to the measure of magic with
matrix product state, and its speedup over standard methods. The speedup
in computation time also leads to a reduced integrated energy consumption of
the CPU as our additional data in Ref.~\cite{Ballarin2024Suppl} shows for one example.

Future directions of this approach for sampling can go into the direction
of comparing parallelized versions, optimizing the caching strategy, or exploring the potential benefit in
other algorithms and applications. While the standard sampling trivially
parallelizes via data parallelism, \opes{} can, for example,
parallelize over the groups of the sorted random numbers. For the application
side, there exists a plethora of candidates with potential speedup
where the examples given within this manuscript give a first intuition which
applications benefit most from \opes{}, i.e., quantum states
with a decaying probability distribution.
Furthermore, since \opes{} only needs the computation of expectation values
of local operators and marginal probabilities it could be applied to any
isometric tensor network such as isoPEPS~\cite{Zalatel2020}. Similarly, the strings of projective measurements can be naturally replaced with Positive Operator-Valued Measures (POVM).

\emph{Code and data availability $-$}
The code implementing the algorithm for both MPS and TTNs is distributed through
the quantum tea leaves~\cite{qtealeaves_0_5_12} python package of \emph{Quantum TEA}, while the code and data to generate the plots
of this work is available via zenodo~\cite{code,code_dataset}. All figures are available together
with metadata under a CC-BY license at~\cite{Ballarin2024Suppl}.

\emph{Acknowledgments $-$}
We thank Nora Reini{\'c} for providing
finite-temperature states in the tree tensor operator representation. We thank Giovanni Cataldi, Marcello Dalmonte, Martina Frau, Marco Rigobello, Ilaria Siloi, Poetri Tarabunga, Marco Tesoro, and Emanuele Tirrito for discussions and feedback.
We acknowledge financial support
from the Italian Ministry of University and Research (MUR) via PRIN2022 project TANQU, and the Departments of Excellence grant 2023-2027 Quantum Frontiers;
from the European Union via
H2020 projects EuRyQa, and
TEXTAROSSA, the QuantERA projects QuantHEP and T-NISQ, and the Quantum Flagship project  Pasquans2, INFN project Quantum,
from the German Federal Ministry of
Education and Research (BMBF) via the funding program quantum
technologies $-$ from basic research to market $-$ project QRydDemo, and
from the World Class Research Infrastructure $-$ Quantum Computing
and Simulation Center (QCSC) of Padova University.
This work was performed in part at the Aspen Center for Physics, which is supported by National Science Foundation grant PHY-2210452; the participation of D. J. at the Aspen Center for Physics was supported by the Simons Foundation.
We also acknowledge computation time supported support by the state of Baden-Württemberg through bwHPC
and the German Research Foundation (DFG) through grant no INST 40/575-1 FUGG (JUSTUS 2 cluster),
the Atos Dibona cluster via TEXTAROSSA, as well
as computation time on Cineca's \emph{Galileo100} machine.

\bibliographystyle{quantum}
\bibliography{refs}

\appendix
\section*{Supplemental material}

The appendices focus on providing additional information which is necessary
to implement the \opes{} algorithm efficiently. Section~\ref{app:real_algorithm}
extends the pedagogical, simplified algorithm to measuring multiple shots
in superiterations, while Sec.~\ref{app:proj_meas_tn} explains the tensor
network aspects and how to cache information for the tensor network.

\section{Implementation of the algorithm}\label{app:real_algorithm}

Section~\ref{sec:methods} contains already one version of the algorithm; now,
we extend the description to cover optional features like multiple shots,
superiterations, or cache access.
The algorithm takes as input the number of random numbers to be sampled $n_r$, the tensor network,
and the set $\Delta=[0,1]\setminus \overline{\Delta} = \{ \delta_{1}, \ldots, \delta_{m} \}$ of $m$ known intervals $\delta_{\alpha}$. A superiteration of the algorithm has the following steps:
\begin{enumerate}
    \item Set the state $\alpha^{*}$ to the empty bitstring, i.e., meaning no measurement has taken place so far. Therefore, the interval is $\Delta=[0, 1]$ and set the initial left boundary of your probability interval $b^l_{\alpha^*}$ and the state probability $p_{\alpha^*}$:
        \begin{align}
            b^l_{\alpha^*}= 0, \quad p_{\alpha^*}=1 \, .
        \end{align}
    \item Sample $n_r$ uniform random variables $\{u_{j}\}_{j=1,\dots,n_r}$ from the interval $\Delta$.
       Sort the random numbers $\{u_{j}\}_{j=1,\dots,n_s}$ in ascending order. This step ensures that the
       cached memory is optimally managed.

   \item Pick the next random number $u_{j}$. If the cache is not empty, find the longest possible
       bitstring matching $u_{j}$ and set $\alpha^{*}$ to the
       longest matching bitstring; at least for the first qubit, there must be a match amongst the
       known intervals. Then, check
       \begin{itemize}
           \item if the cache is empty, continue with the next step.
           \item if the $\alpha^{*}$ contains already all qubits, continue with this step
                 and $u_{j + 1}$, i.e., $u_{j} \in \overline{\Delta}$.
           \item else: load the corresponding tensors from the cache and continue directly with step \emph{5}. 
       \end{itemize}
        
    \item Compute the probability $p_0=p(q_0=\ket{0})$ of measuring $\ket{0}$ on site $0$ and project it on a given state according to its probability and the random number $u_{j}$:
        \begin{align}
            q_0 &= \begin{cases}
            \ket{0} & \text{if } u_{j}<p_0, \\
            \ket{1} & \text{if } u_{j}\geq p_0,
            \end{cases} \\
            \ket{\alpha} &= \ket{\alpha^{*},q_{0}} \, .
        \end{align}
        Update the left boundary and the state probability according to the measure:
        \begin{align}
            b^l_{\alpha} &= \begin{cases}
            b^l_{\alpha^*} & \text{if } u_{j}<p_0, \\
            b^l_{\alpha^*} + p_0 & \text{if } u_{j}\geq p_0
            \end{cases}\\
            p_{\alpha} &= p_{\alpha^*} \cdot p(q_0).
        \end{align}
    
    \item Moving to the next qubit $i$, set $\alpha^{*} = \alpha$; $\alpha^{*}$ is now the bitstring for the qubits up to the $(i-1)^{\mathrm{th}}$ qubit.
        Compute the conditional probability $p_{0|\alpha^{*}}=p(q_i=\ket{0}|\alpha^{*})$.
        Select the measured state according to $u_{j}$ as in the previous point
        and set $b^l_{\alpha}, p_{\alpha}$ accordingly, i.e.:
        \begin{align}
            q_i &= \begin{cases}
            \ket{0} & \text{if } u_{j} < b^l_{\alpha^*} + p_{0|\alpha^{*}}, \\
            \ket{1} & \text{if } u_{j} \geq b^l_{\alpha^*} + p_{0|\alpha^{*}},
            \end{cases} \\            \ket{\alpha} &= \ket{\alpha^{*},q_{i}} \, , \\
            b^l_{\alpha} &= \begin{cases}
            b^l_{\alpha^*} & \text{if } u_{j} < b^l_{\alpha^*} + p_{0|\alpha^{*}} \, , \\
            b^l_{\alpha^*} + p_{0|\alpha^{*}} p_{\alpha^*} & \text{if } u_{j} \geq b^l_{\alpha^*} + p_{0|\alpha^{*}} \, ,
            \end{cases}\\
            p_{\alpha} &= p_{\alpha^*} \cdot p(q_{i}|\alpha^{*}).
        \end{align}

    \item Repeat step \emph{5} until all the sites of the system are measured. Further details for this step are
      explained in Sec.~\ref{sec:methods} and remain valid here.

    \item Remove the measured probability interval from our sampling interval, i.e., $\Delta=\Delta\setminus \delta_{\alpha}$.
        Each new sampled number $u_{j}~\sim U(\Delta)$ explores a 
        different probability interval $\delta_{\beta\neq\alpha}$.
        Update the cache with the corresponding information and tensors.

    \item Continue with step \emph{3} if there are random numbers left for this superiteration.

    \item If the number of superiterations or the desired coverage has been reached, exit. If the next superiteration
    is executed with the current cache, continue with step \emph{2}. For the next superiteration
    with an empty cache, re-start with step \emph{1}.
\end{enumerate}

\subsection{Uniformly sampling from $\Delta$}
In this section, we discuss how the $n_r$ uniform random variables $\{u_j\}_{j=1,\dots,n_r}$ from the interval $\Delta$ are sampled.
First, we define $\Delta_i$ the $M$ continuous intervals componing $\Delta$, i.e.
$\Delta=\bigcup_{i=1}^M \Delta_i$. We define the probability related to the interval $\Delta_i$ as $p_i$.
The technique to sample $u_j$ is as follows:
\begin{enumerate}
    \item Determine how many samples $n_r^i$ are sampled from each interval $\Delta_i$, with $\sum_{i=1}^M n_r^i=n_r$.
    This task can be easily done by sampling $n_r$ numbers uniformly in $U(0,1)$, and identify the sample $u_k\sim U(0,1)$ as belonging to $\Delta_i$ if:
    \begin{align}
      \widetilde{p}_{i-1}\leq u_k < \widetilde{p}_i,
    \end{align}
    with $\widetilde{p}_0=0$ and $\widetilde{p}_i=p_i/\sum_{i=1}^Mp_i$.
    \item Sample $n_r^i$ random numbers from the interval $\Delta_i$:
    \begin{align}
        \{u_j\}_{j=1,\dots,n_r} &= \bigcup_{i=1}^M\; \big\{ u_{j'}\sim U(\Delta_i)\big\}_{j'=1,\dots,n_r^i}\\
        &=\bigcup_{j=1}^{n_r}\; \big\{u_j\sim U(\Delta)\big\}
    \end{align}
\end{enumerate}

The resulting $\{u_j\}_{j=1,\dots,n_r}$ are uniformly distributed over $\Delta$.
We stress that the computational complexity $C_\Delta$ of sampling $n_r$ random numbers from a discontinuous interval $\Delta$ with $M$ continuous intervals $\Delta_i$ scales as:
\begin{align}
    C_\Delta = O\big( n_rM \big),
\end{align}
since we just need to generate $2n_r$ random numbers, but we have to check up to $M$ conditions on $n_r$ of them.

\begin{figure}[!h]
    \centering
    \includegraphics[width=0.5\textwidth]{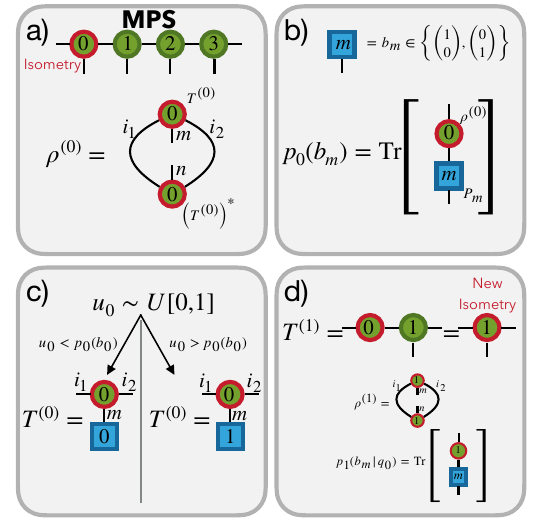}
    \caption{
    \emph{Algorithm to sample from an $n$-sites MPS of local dimension $d=2$ and bond dimension $\chi$ with complexity $O(n\chi^2d^2)$}. Tensors are represented as circles and squares, connected legs are summed over.
    a) Computation of the reduced density matrix $\rho^{(0)}$ using the isometry center of the tensor network.
    b) Computation of the probability of measuring the basis states $b_0=\ket{0}$ and $b_1=\ket{1}$ using the
       projector $\{P_m=\ket{b_m}\bra{b_m}\}_{m=0, 1}$.
    c) Projection of $T^{(0)}$ on $b_0$ with probability $p_0(b_0)$ or on $b_1$ with probability $1-p_0(b_0)$.
    d) Movement of the isometry center from $T^{(0)}$ to $T^{(1)}$, repeating the same computations of a) and b)
    }
    \label{fig:sampling_mps}
\end{figure}

\section{Details of the tensor networks implementation}\label{app:proj_meas_tn}
%
The fundamental requirement for the sampling
is the  ability to efficiently compute expectation values of local observables
and the conditional probabilities of a given quantum state. Here, we review how
these values are computed focusing on the standard sampling procedure,
going afterward into the details of what we cache to optimize the process.

Suppose we want to sample from the $n$-body quantum state $\ket{\psi}$ represented as an
isometric tensor network~\cite{Zalatel2020} with bond dimension $\chi$.
In Fig.~\ref{fig:sampling_mps}, we report the procedure for an MPS with local dimension $d=2$. In general, the steps (a) to (d) following Fig.~\ref{fig:sampling_mps}
are the following:

\begin{enumerate}[label=(\alph*)]
    \item We set the isometry center to the left-most
        physical tensor $T^{(0)}$, in such a way that the contraction over the auxiliary
        legs leads to the single-site reduced density matrix $\rho_{0}$ with:
        \begin{align}
            \rho^{(0)}_{mn} = \sum_{i_1, i_2\dots i_{n}} T^{(0)}_{m,i_1, i_2\dots i_{n}}\left(T^{(0)}_{n,i_1, i_2\dots i_{n}}\right)^*,
        \end{align}
        where $m, n$ represent the physical indexes of local dimension $d$, while the auxiliary indexes, i.e.,
        indexes not representing the Hilbert space of site $0$, are defined as $i_1, i_2\dots i_{n}$.
        The complex conjugate tensor is indicated by a superscript $*$.
    \item Defining the computational basis of a $d$-dimensional qudits as
        $\{b_m\}_{m=0,\dots,d-1}$, we compute the probability of site $0$ being in state $b_m$ as:
        \begin{align}
            p_0(b_m)=\bra{\psi} P_i \ket{\psi} = \mathrm{Tr} \left[ \rho^{(0)} P_{m} \right],
        \end{align}
        where $\left\{P_m=\ket{b_m}\bra{b_m}\right\}_{m=0,\dots,d-1}$ are the projectors over the computational basis.
    \item We randomly select a basis state $b_m$ with probability $p_0(b_m)$ and project the physical tensor $T^{(0)}$ on
        that state:
        \begin{align}
            T^{(0)}_{i_1, i_2\dots i_{n}} = \sum_{m} b_m T^{(0)}_{m, i_1, i_2\dots i_{n}} \, .
        \end{align}
        We call $q_0$ the measured basis state on site $0$.
        The projection is done with a vector and not a projector for efficiency. Effectively,
        this step reduces the local Hilbert space to a one-dimensional Hilbert space and
        reduces the computational complexity.
    \item The isometry center is then moved to the next
        physical tensor $T_1$; by repeating the same procedure, we obtain the conditional probabilities of
        the second site:
        \begin{align}
            p_1(b_m|q_0) =  \mathrm{Tr} \left[ \rho^{(1)} P_{m} \right] \, ,
        \end{align}
         and project into the basis state with these probabilities. Calling $q_1$ the measured basis state on site $1$,
         we denote the probability as $p(q_1|q_0)$.
\end{enumerate}

Iterating through the whole tensor network returns a single sample $\alpha=q_{0},\dots,q_{{n-1}}$ from the quantum state:
\begin{align}
    p_\alpha &= p_{q_{0},\dots,q_{{n-1}}}  \\
    &= p\left(q_{{n-1}}|q_{{n-2}}\dots q_{{0}}\right)\dots  p(q_{1}|q_{0})p(q_{0}). \nonumber
\end{align}
Notice that following this procedure the state $\alpha$ is measured with probability $p_\alpha$, thus this
sampling method is unbiased, e.g., with respect to the order of qubits.

The entire procedure scales as $O(n\chi^2d^2)$ for MPS~\cite{Stoudenmire2010}, while as $O(n \chi^4)$  for TTNs.
Since each sample is independent from the others the standard sampling can be trivially parallelized with OPENMP~\cite{Chandra2001}.

\subsection{Caching the MPS intermediate measurements}
%
\begin{figure}[t]
    \centering
    \vspace{0.5cm}
    \includegraphics[width=0.5\textwidth]{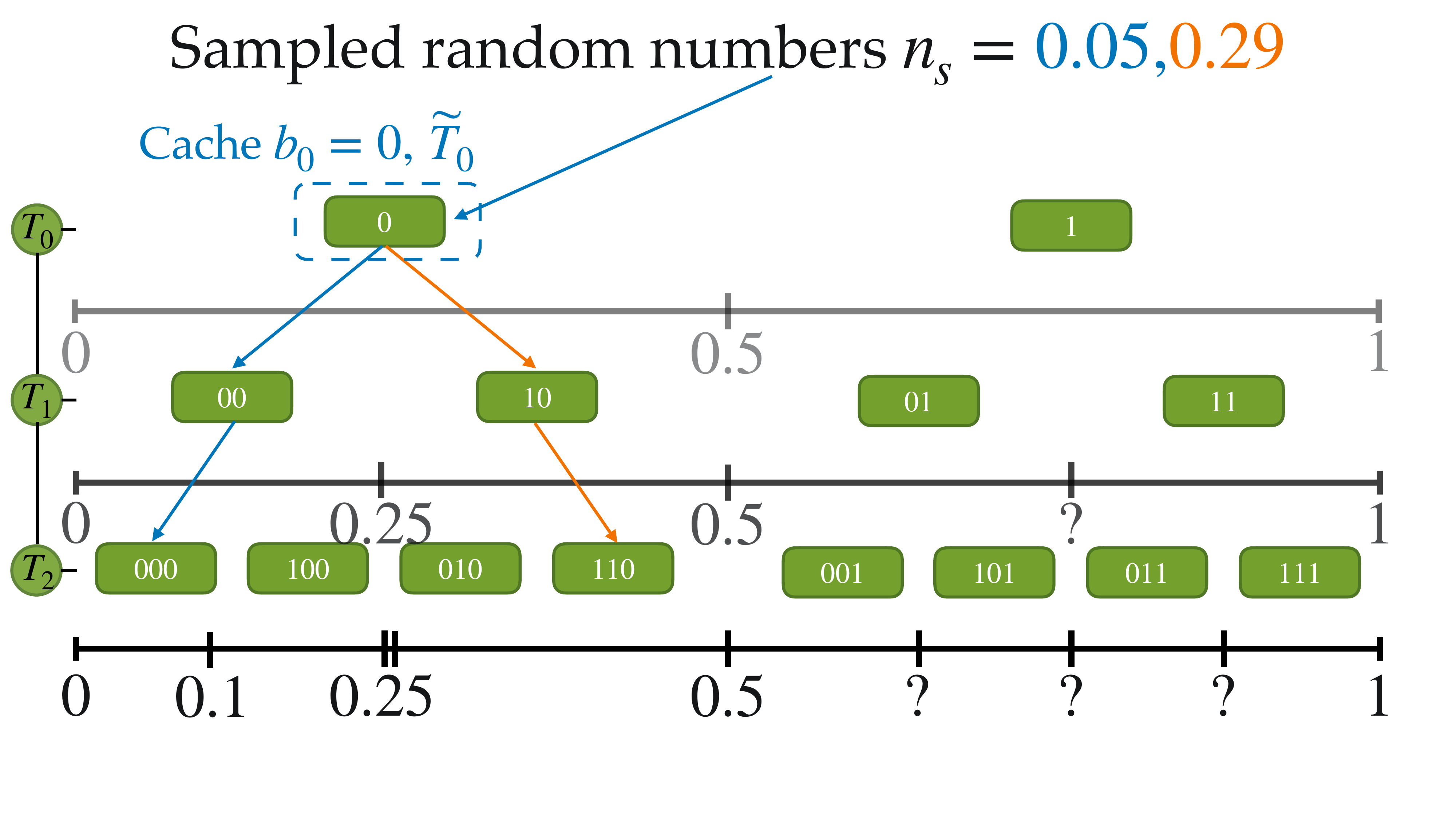}
    \caption{\emph{Caching for matrix product states (MPS).}
        Example of \opes{} with $n=3$ sites and local dimension $d=2$.
        The intervals represent the conditional probabilities of the state represented above it, i.e., the probability
        $p_0(0)=0.5$,  $p_1(0|p_0=0)=0.5$, and so on. First, we compute the probability associated with sample $s_0=0.05$
        (blue line), obtaining the probability interval $\delta_0=(0, 0.1)$ associated to the state $\ket{000}$. We cache the couples
        $(0, \widetilde{T}_0)$, $(00, \widetilde{T}_1)$, where with $\widetilde{T}_i$ we indicate the tensor after the projection with the
        measured basis state. When we continue with the sample $s_1=0.29$, we use the cached information and do not perform any operation on $T_0$.
        At the level of $T_1$, we go to the right instead of the left; thus, we can eliminate the cached information for $T_1, T_2$. We can do so because
        any further samples go to the right of the current interval, due to the ascending ordering.
    }
    \label{fig:mps_caching}
\end{figure}
One of the reasons for the speedup of \opes{} over the standard
sampling is the caching of the intermediate states inside a superiteration.
We suppose the left-to-right sampling ordering discussed in Sec.~\ref{app:proj_meas_tn}, and
that all $\{T_j\}_{0\leq j<m}$, with $m>0$, have been measured.
To restart the computation from tensor $T_m$, it is sufficient
to store the measured state $|b_0,b_1,\dots,b_{m-1}\rangle$ and the tensor $T_m$ contracted
with tensor $T_{m-1}$ after the projection on state $|b_{m-1}\rangle$.
The caching in the superiteration is optimal because, due to the ascending ordering
of the samples $n_s$, we know that once the basis $b_i$ of the tensor $T_i$ pass from
$0$ to $1$ we can delete all the cached information for $j<i$ since we completely explored
that probability interval. The caching structure is shown in Fig.~\ref{fig:mps_caching} for
an MPS with $n=3$ sites and $d=2$ basis. Performance can be further optimized by
keeping the cache between superiterations.

This procedure can be parallelized to using OPENMP~\cite{Chandra2001}, by assigning to different
processors different "branches" of the tree, such as the blue and orange path from tensor $T_1$ in
Fig.~\ref{fig:mps_caching}. Ensuring an equal workload for each process is more challenging
for the \opes{} than for the standard sampling.

\subsection{Caching the TTN intermediate measurements}
%
\begin{figure}
    \centering
    \vspace{0.5cm}
    \includegraphics[width=0.5\textwidth]{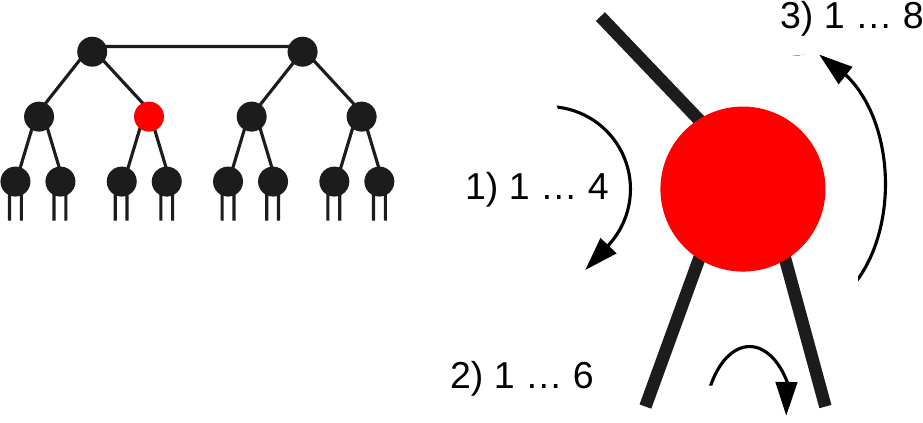}
    \caption{\emph{Caching for tree tensor networks (TTN).}
        The caching for the TTN distinguishes which sites have been
        measured so far. Therein, an auxiliary tensor in the upper layers
        of the TTN can be traversed up to three times. The cache recognizes
        the direction based on the length of the state.
    }
    \label{fig:ttn_caching}
\end{figure}

The caching in the MPS is simplified due to the fact that the MPS is a chain and we respect
the order of the MPS and treat one site after another; no tensor in the MPS is visited twice.
The TTN does not fulfill this assumption and tensors can be visited up to three times during
the projective measurements.

Figure~\ref{fig:ttn_caching} sketches the problem for a 16-site TTN and
one path towards the solution. The tensors that are in the bulk, with
respect to the layers and within a layer itself, are traversed three times with
QR decompositions while running the projective measurements on the children. The tensor
highlighted in red in Fig.~\ref{fig:ttn_caching} is one example. Therefore, the
cache stores tensors for the same position of the TTN multiple times, distinguished
by the length of the bitstring measured so far.


\end{document}